\documentclass[%showpacs,
showkeys,12pt,
preprint,preprintnumbers,nofootinbib,
groupedaddress,superscriptaddress,amsmath,amssymb]{revtex4}
\usepackage{graphicx}% Include figure files
\usepackage{dcolumn}% Align table columns on decimal point
\usepackage{bm}% bold math
\usepackage{amssymb}
\usepackage{amsmath}
\usepackage{epsfig}    
\usepackage{color}
\usepackage{slashed}
\usepackage{hhline}
\def\be{\begin{equation}}
\def\ee{\end{equation}}
\newcommand{\bea}{\begin{eqnarray}}
\newcommand{\eea}{\end{eqnarray}}
\newcommand{\nn}{\nonumber}

\numberwithin{equation}{section}

\begin{document}

%%%%%%%%%
\title{
X-ray line in Radiative Neutrino Model with Global $U(1)$ Symmetry
%Three Loop Neutrino Model and Dark Matters with Global $U(1)'$ Symmetry
}
\preprint{KIAS-P14040}
\author{Hiroshi Okada}
\email{hokada@kias.re.kr}
\affiliation{School of Physics, KIAS, Seoul 130-722, Korea}

\author{Yuta Orikasa}
\email{orikasa@kias.re.kr}
\affiliation{School of Physics, KIAS, Seoul 130-722, Korea}

\date{\today}

\begin{abstract}
We study a three loop induced radiative neutrino model with global $U(1)$ symmetry at TeV scale, in which we consider two component dark matter particles.
We discuss the possibility to explain the X-ray line signal at about 3.55 keV recently reported by XMN-Newton X-ray observatory using data of various galaxy clusters and Andromeda galaxy.
Subsequently, we also discuss to show that sizable muon anomalous magnetic moment, a discrepancy of the effective number of neutrino species $\Delta N_{\rm eff}\approx$ 0.39,
and scattering cross section detected by direct detection searches can be derived.
\end{abstract}
\maketitle
\newpage

\section{Introduction}
Neutrinos and dark matter (DM) physics apparently comes into a new physics beyond the standard model (SM).
One of the elegant scenarios simultaneously to explain them is to generate the neutrino masses at multi-loop level~\cite{Ma:2006km, Aoki:2013gzs, Dasgupta:2013cwa, Krauss:2002px,
Aoki:2008av, Schmidt:2012yg, 
Bouchand:2012dx, Aoki:2011he, Farzan:2012sa,
Bonnet:2012kz, Kumericki:2012bf, Kumericki:2012bh, Ma:2012if, Gil:2012ya,
Okada:2012np, Hehn:2012kz, Dev:2012sg, Kajiyama:2012xg, Okada:2012sp,
Aoki:2010ib, Kanemura:2011vm, Lindner:2011it, Kanemura:2011mw,
Kanemura:2012rj, Gu:2007ug, Gu:2008zf, Gustafsson, Kajiyama:2013zla, Kajiyama:2013rla,
Hernandez:2013dta, Hernandez:2013hea, McDonald:2013hsa, Okada:2013iba,
Baek:2013fsa, Ma:2014cfa, Baek:2014awa, Ahriche:2014xra, Kanemura:2011jj, Kanemura:2013qva, Okada:2014nsa, Kanemura:2014rpa, Chen:2014ska, Ahriche:2014oda, Okada:2014vla, Ahriche:2014cda, Aoki:2014cja, Lindner:2014oea},~\cite{Ahn:2012cg, Ma:2012ez, Kajiyama:2013lja, Kajiyama:2013sza, Ma:2013mga, Ma:2014eka},~\cite{MarchRussell:2009aq}, in which DM could be a messenger particle to tie  the neutrinos to the Higgs boson. Thus we can naturally interpret  the reason why  the neutrino masses are so tiny.
  
 In view of DM, two groups recently reported anomalous X-ray line signal at 3.55 keV from the analysis of XMN-Newton X-ray observatory data of various galaxy clusters and Andromeda galaxy~\cite{Bulbul:2014sua, Boyarsky:2014jta}.
 Once one applies the decaying DM scenario, such a X-ray line can be clearly explained by a 7.1 keV DM mass with small mixing angle; $\sin^22\beta\approx10^{-10}$,  between DM and the neutrinos. 
Since the fact provides a lot of implications on the nature of DM,  many works have been studied~\cite{Ishida:2014dlp, Finkbeiner:2014sja, 
Higaki:2014zua, Jaeckel:2014qea,Lee:2014xua, Kong:2014gea,
 Frandsen:2014lfa, Baek:2014qwa, Cline:2014eaa, Modak:2014vva,
 Babu:2014pxa, Queiroz:2014yna, Demidov:2014hka, Ko:2014xda,
 Allahverdi:2014dqa,  Kolda:2014ppa, Cicoli:2014bfa, Dudas:2014ixa, Choi:2014tva, Okada:2014zea, Chen:2014vna, Conlon:2014xsa, Robinson:2014bma, Liew:2014gia, Chakraborty:2014tma, Tsuyuki:2014aia, Dutta:2014saa, Chiang:2014xra, Geng:2014zqa, Ishida:2014fra,  Baek:2014poa}.
In our letter, we propose a model
 that such a small mixing can be generated at one-loop level, in which 
 the neutrino masses are generated at three-loop level. To realize it, we  introduce a global continuous $U(1)$ symmetry.
As a subsequent result of the additional symmetry, we can also explain the discrepancy of the effective number of neutrino species $\Delta N_{\rm eff}\approx$ 0.39,
which was suggested by Ref.~\cite{Weinberg:2013kea}. 
As the other aspects, sizable muon anomalous magnetic moment and scattering cross section detected by direct detection searches can be derived.
% and can be tested.

This paper is organized as follows.
In Sec.~II, we show our model building including Higgs potential, neutrino masses, and muon anomalous magnetic moment.
In Sec.~III, we analyze DM properties including relic density, X-ray line, and the direct detection with multicomponent scenario. We summarize and conclude in Sec.~VI.
%In appendices, we show the explicit Higgs potential and ...

%\newpage

%%%%%%%%%%%%%%%%%%%%%%%%%%%%%%%%%%%%%
%\section{The Model}
%\subsection{Model setup}
{\it Model setup}

 \begin{widetext}
\begin{center} 
\begin{table}[b]
%\begin{tiny}
\begin{tabular}{|c||c|c|c|c|c|c||c|c|c|c|c|}\hline\hline  
&\multicolumn{6}{c||}{Lepton Fields} & \multicolumn{5}{c|}{Scalar Fields} \\\hline
& ~$L_L$~ & ~$e_R^{}$~ & ~$E_L$~ & ~$E_R$~ & ~$N_R$& ~$X_R$~ & ~$\Phi$~ &
 ~$\eta$~ & ~$\chi^+_1$~  & ~$\chi^{+}_2$~ & ~$\chi_0$ \\\hline 
$SU(2)_L$ & $\bm{2}$ & $\bm{1}$& $\bm{1}$ & $\bm{1}$ & $\bm{1}$ & $\bm{1}$ &
 $\bm{2}$&$\bm{2}$&$\bm{1}$ &$\bm{1}$  &$\bm{1}$\\\hline 
$U(1)_Y$ & $-1/2$ & $-1$ & $-1$ & $-1$  & $0$  & $0$ & $+1/2$& $+1/2$& $+1$ & $+1$  & $0$  \\\hline
$U(1)$ & $-3x/2$ & $-3x/2$ & $-3x/2$ & $-3x/2$  & $-x/2$   & $-x/2$ & $0$& $0$& $x$ & $x$  & $x$  \\\hline
%%%
$\mathbb{Z}_2$ & $+$ & $+$ & $-$ & $-$  & $-$  & $+$ & $+$& $-$& $-$& $+$ & $+$  \\\hline\hline
\end{tabular}
\caption{Contents of lepton and scalar fields
and their charge assignment under $SU(2)_L\times U(1)_Y\times U(1)\times\mathbb{Z}_2$.}
\label{tab:1}
% \end{tiny}
\end{table}
\end{center}
\end{widetext}

We discuss a three-loop induced radiative neutrino model. 
The particle contents and their charges are shown in Tab.~\ref{tab:1}. 
We add gauge singlet
charged fermions $E_L$ and $E_R$, {three} gauge singlet Majorana fermions $N_R$, and a gauge singlet Majorana DM $X_R$.
For new bosons, we introduce $SU(2)_L$ doublet scalars $\eta$, two
singly-charged  singlet scalars ($\chi^+_1, \chi_2^{+}$), and a neutral singlet scalar $\chi_0$ to the SM.
%%% 
We assume that  only the SM-like Higgs $\Phi$ and $\chi_0$ have vacuum
expectation values (VEVs), which are symbolized by $v$ and $v'$ respectively. 
We also introduce a global $U(1)$ symmetry, under which $\Phi$ and $\eta$ do not have the charge in order not to couple to the goldstone boson (GB)~\cite{Baek:2014awa}.
$x\neq0$ is an arbitrary number of the charge of $U(1)$ symmetry, and their assignments can realize our neutrino model at three loop level.
%Otherwise the $\mathbb{Z}_2$ symmetry which guarantees DM stability is spontaneously broken. 
The $Z_2$ symmetry assures the stability of DM that is the neutral component of $\eta$.

The renormalizable Lagrangian for Yukawa sector, mass term, and scalar potential
under these assignments are given by
\begin{align}
\mathcal{L}_{Y}
&=
y_\ell \bar L_L \Phi e_R  + y_{\eta} \bar L_L \eta E_R  %+ y_{R}  \bar E^c_R N_R \chi^{+}_2
+y_{\chi_1}\bar E_L X_R \chi^{-}_1 +y_{\chi_2}\bar E_L N_R \chi^{-}_2  
\nn\\
&+ y_N \chi_0 \bar N^c_R N_R + y_X \chi_0 \bar X^c_R X_R + M_E\bar E_L E_R+\rm{h.c.} \\ 
%%%
\mathcal{V}
&= 
 m_\Phi^2 |\Phi|^2 + m_\eta^2 |\eta|^2 + m_{\chi_1}^2 |\chi^+_1|^2  + m_{\chi_2}^2 |\chi_2^{+}|^2  +  m_{\chi_0}^2 |\chi_0|^2
 \nn\\
&+ \Bigl[
 \lambda_0 \Phi^T (i\tau_2) \eta \chi^-_1\chi_0 + \lambda_0' (\chi^+_1 \chi^-_2)^2 + {\rm h.c.}\Bigr]
 \nn\\
&
  +\lambda_1 |\Phi|^{4} 
  %%%
  + \lambda_2 |\eta|^{4} 
 %%%
  + \lambda_3 |\Phi|^2|\eta|^2 
  + \lambda_4 (\Phi^\dagger \eta)(\eta^\dagger \Phi)
  \nn\\
& +\Bigl[\lambda_5(\Phi^\dag\eta)^2+{\rm h.c.}\Bigr]
 +\lambda_6  |\Phi|^2|\chi^+_1|^2  %+\lambda_6'  |\Phi|^2|\chi^+_2|^2
+ \lambda_7  |\eta|^2|\chi^+_1|^2
 \nn\\
&+
\lambda_8  |\Phi|^2|\chi_2^{+}|^2 + \lambda_{9}  |\eta|^2|\chi_2^{+}|^2
+
 \lambda_{10} |\chi^{+}_1|^4  +
 \lambda_{11} |\chi_2^{+}|^4
  \nn\\
&+
 \lambda_{12} |\chi^{+}_1|^2|\chi_2^{+}|^2 
 + \lambda_{13}  |\Phi|^2 |\chi_0|^2
+
 \lambda_{14} |\eta|^2 |\chi_0|^2 
  \nn\\
&+
 \lambda_{15}  |\chi_1^+|^2 |\chi_0|^2 +
 \lambda_{16}  |\chi^+_2|^2 |\chi_0|^2+
 \lambda_{17}   |\chi_0|^4
  \nn\\
&
 +\left[\lambda_{18}(\Phi^\dag \eta)(\chi^+_1\chi^-_2)+{\rm h.c.}  \right]
,
\label{HP}
\end{align}
where the first term of $\mathcal{L}_{Y}$ can generates the SM
charged-lepton masses, and we assume 
$\lambda_0$, $\lambda_0'$
%$\mu(\equiv \lambda_0 v'/\sqrt2) $, $\mu_\kappa$
, $\lambda_5$, and $\lambda_{18}$ to be real.
%can be chosen to be real without any loss of generality by renormalizing the phases to scalar bosons. 

%%%
The scalar fields can be parameterized as 
\begin{align}
%\begin{tiny}
&\Phi =\left[
\begin{array}{c}
w^+\\
\frac{v+\phi+iz}{\sqrt2}
\end{array}\right],\quad 
\eta =\left[
\begin{array}{c}
\eta^+\\
\frac{\eta_{R}^{}+i\eta_{I}^{}}{\sqrt2}
\end{array}\right],\
\chi_0=\frac{v'+\sigma}{\sqrt{2}}e^{iG/v'}
.   
\label{component}
%\end{tiny}
\end{align}
where $v~\simeq 246$ GeV is the VEV of the Higgs doublet, and $w^\pm$
and $z$ are respectively GB 
which are absorbed by the longitudinal component of $W$ and $Z$ bosons.
Inserting the tadpole conditions; $\partial\mathcal{V}/\partial\phi|_{v}=0$ and $\partial\mathcal{V}/\partial\sigma|_{v'}=0$,
The resulting mass matrix of the CP even boson $(\phi,\sigma)$ 
 is given by
\begin{eqnarray}
&&
m^{2} (\phi,\sigma) = \left[%
\begin{array}{cc}
  2\lambda_1v^2 & \lambda_{13}vv' \\
  \lambda_{13}vv' & 2\lambda_{17}v'^2 \\
\end{array}%
\right] \\
&&= \left[\begin{array}{cc} \cos\alpha & \sin\alpha \\ -\sin\alpha & \cos\alpha \end{array}\right]
\left[\begin{array}{cc} m^2_{h} & 0 \\ 0 & m^2_{H}  \end{array}\right]
\left[\begin{array}{cc} \cos\alpha & -\sin\alpha \\ \sin\alpha &
      \cos\alpha \end{array}\right], \nn
\end{eqnarray}
where $h$ is the SM-like Higgs and $H$ is an additional CP-even Higgs mass
eigenstate. The mixing angle $\alpha$ is given by 
%$\sin 2\alpha=\frac{2\lambda_{13} v v'}{m^2_h-m_H^2}$.
\be
%\tan 2\alpha=\frac{\lambda_{13} v v'}{\lambda_{17} v'^2-\lambda_1 v^2}.
\sin 2\alpha=\frac{2\lambda_{13} v v'}{m^2_h-m_H^2}.
\ee
The Higgs bosons $\phi$ and $\sigma$ are rewritten in terms of the mass eigenstates $h$ and $H$ as
%\begin{eqnarray}
$\phi = h\cos\alpha + H\sin\alpha$, %\nn\\
$\sigma =- h\sin\alpha + H\cos\alpha$.
%\label{eq:mass_weak}
%\end{eqnarray}
GB appears due to the spontaneous symmetry breaking of
the global $U(1)$ symmetry. 

The mass matrix  $M_+^2$ of the singly-charged scalar boson $(\eta^\pm,\chi^\pm_1)$ is given by 
\begin{align}
M_+^2  \!\! =\!\!  \left[%
\begin{array}{cc}
\!\! \!\! m^2_{\eta}+\frac{\lambda_3 v^2+\lambda_{14}v'^2}{2} \!\! \!\!  &\!\!  \!\! \frac{\lambda_0 vv'}{2} \!\! \!\!  \\
\!\! \!\!  \frac{\lambda_0 vv'}{2} \!\! \!\!  &\!\! \!\!  m^2_{\chi_1} + \frac{\lambda_6 v^2+\lambda_{15}v'^2}{2}\!\! \!\!  \\
\end{array}%
\right]. 
\end{align}
The mass eigenstates $h^\pm$, $H^\pm$ are defined by introducing the
mixing angle $\theta$ as 
\begin{align}
\left(\begin{array}{c}
\eta^\pm \\
\chi^\pm_1 
\end{array}\right)=
\left(\begin{array}{cc}
\cos\theta  & \sin\theta\\
-\sin\theta & \cos\theta 
\end{array}\right)
\left(\begin{array}{c}
h^\pm \\
H^\pm 
\end{array}\right).
\end{align}
where the mixing angle $\theta$ is given by 
\be
\sin 2\theta=\frac{\lambda_0 vv' }{m_{h^+}^2-m_{H^+}^2}\label{mix:theta}.
\ee

The other mass eigenstates are given as
\begin{eqnarray}
m^{2}_{\chi^{\pm}_2} &=& m_{\chi_2}^{2}  + \frac12 (\lambda_8 v^{2}+\lambda_{16} v'^{2}), \\ 
 %%%
m^2_{\eta_{R}} &=& 
m_\eta^{2} + \frac12\lambda_{14}v'^2
+ \frac12\left( \lambda_3  + \lambda_4 + 2\lambda_5\right)v^2 , \\  
m^2_{\eta_{I}} &=& 
m_\eta^{2} + \frac12\lambda_{14}v'^2
 + \frac12\left( \lambda_3  + \lambda_4 - 2\lambda_5\right)v^2.
\end{eqnarray}

%\subsection{Neutrino mass matrix}
{\it Neutrino mass matrix}

%%%%%%%%%%%%%%%%%%%
\begin{figure}[t]
\begin{center}
\includegraphics[scale=0.5]{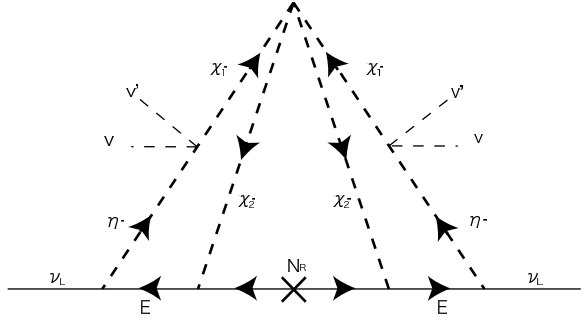}
   \caption{Radiative generation of neutrino masses.}
   \label{neutrino-diag}
\end{center}
\end{figure}
%%%%%%%%%%%%%%%%%%%
The Majorana neutrino mass matrix $m_\nu$ is derived at three-loop level from the
diagrams depicted in Fig.~\ref{neutrino-diag}, which is given by 
%\begin{widetext}
\begin{align}
&(m_{\nu})_{ij}
=
\frac{\lambda_0'}{4(4\pi)^6} %\sum_{\alpha=1}^{3} 
\sum_{\beta,\gamma=1}^{3} %\sum_{\gamma=1}^{2}
% \sum_{\delta=1}^{3}
\left[
 %\frac{
 (y_\eta)_{i}%M_{E^\alpha}
 (y_{\chi_2})_{i\beta}
M_{N_{\beta}} (y_{\chi_2}^T)_{\beta\gamma}
 %(M_{E^\delta})
 (y_{\eta})^T_{j}%}
 %{M_{E^\delta}^2}
 \right] 
% \sin^2\theta \cos^2\theta
 \sin^22\theta
%\nn\\
%%%&\times 
\nn\\
&\left[
F_1\left(
\frac{m_{h^+}^2}{M_{E}^2},
\frac{m_{h^+}^2}{M_{E}^2}
\right)
+
F_1\left(
\frac{m_{H^+}^2}{M_{E}^2},
\frac{m_{H^+}^2}{M_{E}^2}
\right)
%\right.\nonumber\\&\left.\qquad
-2
F_1\left(
\frac{m_{h^+}^2}{M_{E}^2},
\frac{m_{H^+}^2}{M_{E}^2}
\right)
\right],
\end{align}%\end{widetext}
where $M_N\equiv y_Nv'/\sqrt2$ {is assumed to be diagonal, and}  the loop function $F_1$ is computed as
%\begin{widetext}
\begin{align} 
&F_1\left(X_1,X_2\right) 
=
\int d^3x\frac{\delta(x+y+z-1)}{z(z-1)}\nn\\&
\times
\int d^3x'\frac{\delta(x'+y'+z'-1)}{z'(z'-1)}%
\int d^3x''\frac{\delta(x''+y''+z''-1)}
{x''+z'' X_1-y'' \Delta(X_2)},
\end{align}
%\end{widetext}
with 
\begin{align}
&\Delta(X_2)=\frac{y' \frac{M_{N^{\beta\gamma}}^2}{M_{E}^2} + z'  \frac{m_{\chi_2^+}^2}{M_{E}^2} 
-x' \Delta'(X_2)}{z'(z'-1)},\nn\\
& \Delta'(X_2)=
\frac{x  + z  \frac{m_{\chi_2^+}^2}{M_{E^{}}^2} 
-y  X_2}
{z(z-1)},
\end{align}
where we define $d^3x\equiv dxdydz$,  $d^3x'\equiv dx'dy'dz'$, and  $d^3x''\equiv dx''dy''dz''$. 
%One finds rather wide allowed range 
To obtain the neutrino masses reported by Planck data ~\cite{Ade:2013lta}; $m_{\nu}<0.933~\mathrm{eV}$, 
the following is required
\begin{widetext}
\begin{align}
%\frac{
\lambda_0'
%}{(4\pi)^6} \sum_{\alpha=1}^{3} 
%\sum_{\beta=1}^{2} \sum_{\gamma=1}^{2}
% \sum_{\delta=1}^{3}\left[
 %\frac{
% (y_\eta)_{i}%M_{E^\alpha}
 y_{\chi_2}^2%_{\beta}
M_{N^{}} 
%(y_{\chi})_{\gamma}
 %(M_{E^\delta})
% (y_{\eta})_{j}%}
 %{M_{E^\delta}^2} \right] 
% \sin^2\theta \cos^2\theta
%\nn\\&\times 
\left[
F_1\left(
\frac{m_{h^+}^2}{M_{E}^2},
\frac{m_{h^+}^2}{M_{E}^2}
\right)
+
F_1\left(
\frac{m_{H^+}^2}{M_{E}^2},
\frac{m_{H^+}^2}{M_{E}^2}
\right)
%\right.\nonumber\\&\left.\qquad
-2
F_1\left(
\frac{m_{h^+}^2}{M_{E}^2},
\frac{m_{H^+}^2}{M_{E}^2}
\right)
\right]
<
1.17\ {\rm MeV}
,
\end{align}
\end{widetext}
where we fix $\theta=\pi/4$ for simplicity, and $y_\eta^2\approx 4\pi$ to obtain the sizable muon anomalous magnetic moment as discussed   {in the next subsection.
The observed mixing matrix, that is PMNS(Pontecorvo- Maki-Nakagawa-Sakata) matrix ($U_{\rm PMNS}$)~\cite{Maki:1962mu}, can be realized by introducing the Casas-Ibarra parametrization~\cite{Casas:2001sr, Okada:2012fs}.
In our case, the Dirac type Yukawa parameters can be written by
\begin{widetext}
\begin{align}
(y_\eta y_{\chi_2})_{i\beta}
&=U_{\rm PMNS}^* 
%%%
\left(\begin{array}{ccc}
m_{\nu_1}^{1/2} & 0 &0 \\ 0 & m_{\nu_2}^{1/2} &0 \\ 0 & 0 & m_{\nu_3}^{1/2} \\ 
\end{array}\right)
{\cal O} 
\left(\begin{array}{ccc}
M_{N_1}^{-1/2} & 0 &0 \\ 0 & M_{N_2}^{-1/2} &0 \\ 0 & 0 & m_{M_3}^{-1/2} \\ 
\end{array}\right) R^{-1/2},\\
R&=
\frac{\lambda_0'}{4(4\pi)^6}  \sin^22\theta
%%%
\left[
F_1\left(
\frac{m_{h^+}^2}{M_{E}^2},
\frac{m_{h^+}^2}{M_{E}^2}
\right)
+
F_1\left(
\frac{m_{H^+}^2}{M_{E}^2},
\frac{m_{H^+}^2}{M_{E}^2}
\right)
%\right.\nonumber\\&\left.\qquad
-2
F_1\left(
\frac{m_{h^+}^2}{M_{E}^2},
\frac{m_{H^+}^2}{M_{E}^2}
\right)
\right],
%%%
\end{align}
\end{widetext}
 %%%
 where ${\cal O}$ is an arbitrary orthogonal matrix with complex values.
 Then one finds that the neutrino mass eigenvalues $(m_{\nu_1},m_{\nu_2},m_{\nu_3})$ are written by
 \begin{align}
m^\dag_\nu m_\nu
&=U_{\rm PMNS}
%%%
\left(\begin{array}{ccc}
m_{\nu_1}^{2} & 0 &0 \\ 0 & m_{\nu_2}^{2} &0 \\ 0 & 0 & m_{\nu_3}^{2} \\ 
\end{array}\right)
U_{\rm PMNS}^\dag.
\end{align}
%%%%%%%%%%%%%%%%%%%
}

%\subsection{% Lepton Flavor Violations (LFVs) and muon anomalous magnetic moment}
{\it% Lepton Flavor Violations (LFVs) and 
muon anomalous magnetic moment}
%%%%%%%%%%%%%%%%%%%
\begin{figure}[t]
\begin{center}
\includegraphics[scale=0.5]{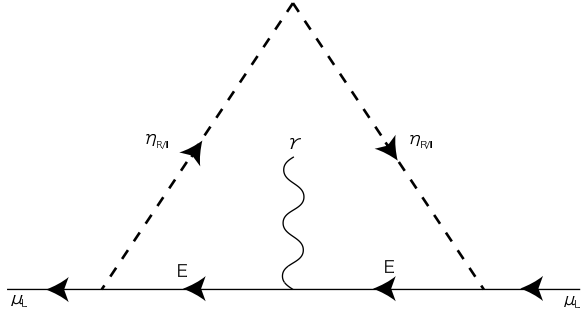}
   \caption{Diagram of the muon anomalous magnetic moment.}
   \label{fig:g-2}
\end{center}
\end{figure}
%%%%%%%%%%%%%%%%%%%
In principle, we obtain the LFV process from the terms which are proportional to $y_{\eta}$. Especially, 
$\mu\to e\gamma$ process  gives the most stringent bound. However since we can fix $y_{\eta}$ to be the diagonal matrix~\footnote{We expect that the mixing of MNS can be obtained by $y_{\chi_2}$, {as discussed in the previous subsection.}}, we can simply avoid such kind of processes.
So we move on to the discussion of the muon anomalous magnetic moment.

The muon anomalous magnetic moment has been 
measured at Brookhaven National Laboratory. 
The current average of the experimental results is given by~\cite{bennett}
\begin{align}
a^{\rm exp}_{\mu}=11 659 208.0(6.3)\times 10^{-10},\notag
\end{align}
which has a discrepancy from the SM prediction by $3.2\sigma$~\cite{discrepancy1} to $4.1\sigma$~\cite{discrepancy2} as
\begin{align}
\Delta a_{\mu}=a^{\rm exp}_{\mu}-a^{\rm SM}_{\mu}=(29.0 \pm
9.0\ {\rm to}\ 33.5 \pm
8.2)\times 10^{-10}. \label{g-2_dev}
\end{align}
We have a contribution on the this process through the term of $y_\eta$, as can be seen in Fig.~\ref{fig:g-2}.
The formula is given as
\begin{align}
\Delta a_\mu &= \frac{1}{2 (4\pi)^2}%\sum_{\alpha=1}^3
|y_{\eta}^{}|^2\left(\frac{m_\mu}{M_{E}}\right)^2\left[
F_2\left(\frac{m_{\eta_R}^2}{M_{E}^2}\right)
+F_2\left(\frac{m_{\eta_I}^2}{M_{E}^2}\right)\right],
\label{eq:muon-g-2}
\end{align}
where
\begin{align}
F_2(x)&= \frac{1-6x+3x^2+2x^3-6x^2\ln x}{6(1-x)^4}.
\end{align}
%%%
We can obtain sizable muon anomalous magnetic moment
\begin{align}
\Delta a_\mu\approx 1.5\times 10^{-9},
\end{align}
 if we set $y_\eta\approx{\cal O}(\sqrt{4\pi})$ that is limit of the perturbative 
which is  within the 2$\sigma$ error,theory, $M_E\approx{\cal O}(300)$ GeV that comes from the analogy of the slepton search of LHC~\cite{altas}, and $m_{\eta_R}\approx m_{\eta_I}=67.83$ GeV, which can be obtained from the DM analysis as can be seen in the next section.

%\newpage

%\section{Dark Matter Particles}
{\it Dark Matter Particles}
%\label{sec:DM}
We have two DM candidates $X_R$ and $ \eta_{R}$, which do not interact each other at tree level.
Hence two component scenario can be taken in consideration. Since $X_R$ can be expected to explain the X-ray line at 3.55 keV,
its mass $M_X\equiv y_Xv'/\sqrt2$ be 7.1 keV.  On the other hand, since $\eta_R$ is expected to be detected  direct detection searches such as LUX~\cite{Akerib:2013tjd}, its mass range be $m_{\eta_R}\approx\cal{O}$(10-80) GeV~\cite{Hambye:2009pw}, where we restrict ourselves the mass be less than the mass of the SM gauge bosons to forbid the too large cross section.
Hereafter we simply assume that the number density of DMs is the same rate, that is, $\Omega_{X_R} h^2: \Omega_{\eta_R} h^2=1:1$.
Also we suppose that both are assumed to be the cold DMs, and
the mixing of $\alpha$ sets to be zero to analyze the cross section of the relic density because it is not so sensitive to the cross section.

%\subsection{$X_R$ dark matter}
{\it $X_R$ dark matter}

%%%%%%%%%%%%%%%%%%%
\begin{figure}[t]
\begin{center}
\includegraphics[scale=0.5]{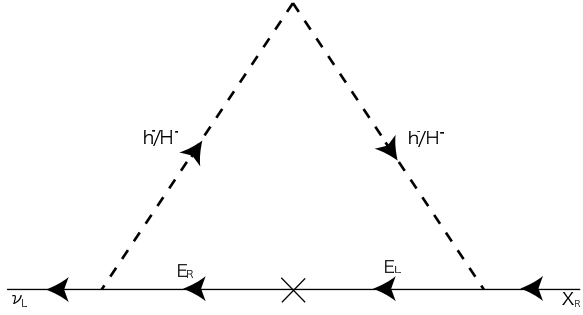}
   \caption{Mixing between neutrinos and DM.}
   \label{fig:mix}
\end{center}
\end{figure}
%%%%%%%%%%%%%%%%%%%
The dominant relativistic cross section of $X_R$, which is $2X_R\to H\to2G$ via s-channel,  is given by
\bea
(\sigma v)_{\rm rel}\approx \frac{M_X^6}{4\pi v'^4}\frac{v^2_{\rm rel}}{(4M_X^2-m_H^2)^2}.
\eea
%%%
 To obtain the correct relic density $\Omega_{X_R} h^2=0.12/2$~\cite{Ade:2013lta}, the  required cross section be
 \bea
 (\sigma v)_{\rm rel}\approx 3.06\times 10^{-8}\ {\rm GeV^{-2}}.
 \eea
Once we set $v'\approx1$ GeV, $m_H$ be 
\bea
m_H\approx 1.421\times 10^{-5}\ {\rm GeV}\approx 2 M_X.
\eea
The above result implies that a mild fine-tuning is needed~\footnote{Since the decay rate of the $H$ is very tiny as well as the one of $h$, we neglect these contributions.}.

Here we consider the contribution of GB to the effective number of neutrino species $\Delta N_{\rm eff}\approx$0.39 suggested by~\cite{Weinberg:2013kea}.
It can be realized when the appropriate era of
freeze-out of the Goldstone boson is before muon annihilation while the other SM particles
are decoupled. Thus it corresponds to $T\approx m_\mu$, where $T$ is the temperature of the universe.
The scattering of the Goldstone boson with the SM particles occurs through the Higgs
exchange. Then the interaction rate be the same order as the Hubble parameter at $T\approx m_\mu$ .
%To obtain such a observed value,
Considering the above process, one leads to the following relation~\cite{Baek:2013fsa},
\bea
\sin2\alpha\approx\sqrt{\frac{4(vv')^2(m_hm_H)^4}{(m_h^2-m_H^2)^2m^7_\mu m_{pl}}}\approx7.33\times 10^{-14},
\eea
where $m_{pl}\approx1.22\times10^{19}$ GeV, and $m_\mu\approx0.106$ GeV.  
%It suggests that the mixing of $\alpha$ is very tiny.

The mixing between $X_R$ and the active neutrinos can be obtained at one-loop level as depicted in Fig~\ref{fig:mix}, and it is given by
\bea
\beta_{X_R-\nu}&=&\frac{M_E y_{\chi_1} y_\eta}{2\sqrt2(4\pi)^2M_X}\int^1_0dx\ln\left[\frac{x+(1-x)\Delta_{h^+}}{x+(1-x)\Delta_{H^+}}\right]
\nn\\
&\approx& 7.1\times 10^{-6},
\eea
where $\Delta_{h^+}\equiv m^2_{h^+}/M^2_E$ and $\Delta_{H^+}\equiv m^2_{H^+}/M^2_E$.
When $\Delta_{h^+}$ and  $\Delta_{H^+}$ are larger than 1 and $y_\eta y_{\chi_1}\approx 0.001$, 
we obtain 
\begin{equation}
\lambda_0 = \frac{m^2_{H^+}-m^2_{h^+}}{vv'}\approx 1.91\times10^{-6} \left(\frac{M_E}{{\rm GeV}}\right)^2
%\nn\\
\approx 0.172,
\end{equation}
combining with Eq.~(\ref{mix:theta}).
%, we obtain \bea\lambda_0\approx 0.172.\eea

%\subsection{$\eta_R$ dark matter}
{\it $\eta_R$ dark matter}

The dominant relativistic cross section of $\eta_R$, which is ${\eta_R}\to H\to2G$ via s-channel and  ${\eta_R}\to2H$\footnote{The $t$- and $u$-processes through the term of $y_\eta$ can be negligible because of the  $d$-wave suppression.}
,
is given by
\bea
(\sigma v)_{\rm rel}\approx \frac{5\lambda_{14}^2}{64\pi m_{\eta_R}^2}.
\eea
%%%
 To obtain our relic density $\Omega_{\eta_R} h^2=0.12/2$, the  required cross section be
 \bea
 (\sigma v)_{\rm rel}\approx 5.41\times 10^{-9}\ {\rm GeV^{-2}}.
 \eea
Once we set $\lambda_{14}^2\approx0.001$, we obtain%$m_{\eta_R}$ be 
\bea
m_{\eta_R}\approx 64.38\ {\rm GeV} .
\eea
%The above result implies that a mild fine-tuning is needed~\footnote{Since the decay rate of the $H$ is very tiny as well as the one of $h$, we neglect the contribution.}.

%\subsection{Direct Detection}
{\it Direct Detection}:

$\eta_R$ can be tested  by the spin independent elastic scattering cross section, and it form is give by
\bea
&&\sigma\approx \frac{0.079}{\pi}\times
 \\&&
\left[\frac{m_p}{m_{\eta_R}v}\right]^2
\left[\frac{(\lambda_3+\lambda_4+2\lambda_5)v}{m^2_h}\cos\alpha+ \frac{\lambda_{14}v'}{m^2_H}\sin\alpha
\right]^2 ({\rm GeV})^2,\nn
\eea 
where $m_p\approx$0.938 GeV is the proton mass.
The current lowest bound of $\sigma$ can be found by the experiment of LUX, which be around ${\cal O}(10^{-45})$ cm$^2$ at $m_{\eta_R}\approx64$ GeV.
Inserting all the fixed parameters, we can satisfy this constraint if 
\bea
\lambda_3+\lambda_4+2\lambda_5\le0.011.
\eea

%\section{Conclusions}
{\it Conclusions}:

We have constructed a three-loop induced neutrino model with a global $U(1)$ symmetry, in which we have naturally
explained the X-ray line signal at about 3.55 keV with $X_R$ recently reported by XMN-Newton X-ray observatory using data of various galaxy clusters and Andromeda galaxy.
Subsequently, we have also shown that sizable muon anomalous magnetic moment within 2$\sigma$ error, which is around $1.5\times 10^{-9}$.
The effective number of neutrino species $N_{\rm eff}\approx$ 0.39 can be derived at an appropriate parameter set due to the additional symmetry.
Since our model has two DM candidates that does not interact each other due to the $U(1)$ symmetry,
another candidate ($\eta_R$) can be tested by the current direct detection searches such as LUX.

%\newpage
%%%%%%%%%%%%%%%%%%%%%%%%%%%%%%%%%%%
\vspace{0.3cm}
%\hspace{0.2cm} {\bf Acknowledgments}
%\section*{Acknowledgments}:
%\vspace{0.5cm}
{\it Acknowledgments}:

H.O. thanks to Prof. Seungwon Baek, Dr. Hiroyuki Ishida, Dr. Eibun Senaha, Dr. Takashi Toma, and Dr. Kei Yaygu for fruitful discussions.
%%%%%%%%%%%%%%%%%%%%%%%%%%%%%%%%%%%

\end{document}